\input phyzzx
\def \sign{ \mathop{ \rm sign}\nolimits}
\def \trace{ \mathop{ \rm trace}\nolimits}
%

DFTUZ /94-11
\vfill
\title{INTEGRABLE MODELS ASSOCIATED TO CLASSICAL REPRESENTATIONS OF  $
{U}_{q}(\widehat{sl(n)}) $}
\author{J. ABAD and M. RIOS}
\address{Departamento de F\'{\i}sica Te\'{o}rica, Facultad de Ciencias,
Universidad de Zaragoza, 50009 Zaragoza, Spain}
\bigskip
\bigskip
\bigskip
\bigskip
\abstract

We describe a representation for $ {U}_{q}(\widehat{sl(n)})$, when $q$ is not a
root of unity, based on the fundamental representation of  $sl(n)$. As
${U}_{q}(sl(n))$ has a Hopf algebra structure with a non-commutative
co-product, we look for a intertwine matrix $R$ that relates two possible
definitions of that co-product. We solve cases for $n=2$ and $n=3$, and then we
generalize for any $n$. We obtain the hamiltonian associated to such matrix
$R$, corresponding to  a  multi-state chain. As the case for $n=2$ corresponds
to the XXZ model with spin $1/2$, for $n>2$ we have the generalization of the
XXZ model to $sl(n)$. We show the case for $n=3$ and its solution by Bethe
ansatz.

\vfill
\eject
\chapter{Introduction}

In these last years, the search for integrable models and related problems has
deserved great attention. Some of the oldest and still most interesting models
are the isotropic and anisotropic spin $1/2$  chains of Heisenberg ( XXX and
XXZ models). The mathematical structure arising in these relatively simple
models is astonishingly rich. The key featuring for it is the existence of a
complete set of mutually commuting integrals of motion. They are usually
defined by means of transfer matrix $F(u)$, which for these models is a
function of a single spectral parameter $u$. Sutherland [1] showed that $[ H ,
F(u) ] = 0$ and later Baxter [2], in a more general context, proved the key
property of the transfer matrix $[F(u), F(v)] =  0 $. This implies that the
logarithmic derivatives of $F(u)$,
$$
{Q}_{n}={d^{n} \over du^{n}}{\log{F(u)}}|_{u={u}_{0}}  \eqno (1.1)
$$
mutually commute. The hamiltonian is related usually to ${Q}_{1}$.

With these results, Faddeev and collaborators stated the main ideas of quantum
inverse scattering method (QISM) [3,4], where it is introduced the monodromy
matrix $T(u)$ acting on an auxiliary space and whose elements are operators in
the tensorial product of the site spaces. The trace of that matrix is transfer
matrix $F(u)$.
The main property of the monodromy matrix is that it verifies the Yan-Baxter
equation (YBE), that resume the commutation relations of before. Besides it is
possible to solve the model by using the method known as algebraic Bethe
ansatz.

The development of quantum groups has introduced the YBE in a consistent
mathematical structure and has guided the search of new solvable spin chains.
It is know that these are related very closely to the representations of the
quantum groups [5]. In fact, the XXZ model is associated to the dimension two
representation of  ${U}_{q}\bigl(sl(2)\bigr) $ when $q$ is not a root of the
unity. Other solvable models have been found for representations of   ${U}_{q}
\bigl(sl (2)\bigr) $ when $q$ is a root of unity [6,7,8] and recently also for
${SU}_{p,q}(2)$ [9].

The present paper is organized  as follows. In the next section, following the
line used for the representations of ${U}_{q}\bigl(\widehat{sl(2)}\bigr) $, we
build the representations of ${U}_{q}\bigl(\widehat{sl(n)}\bigr) $ when q is
not a root of the unity [5,10], and find the $R$ intertwine matrix in the
product of representations. By using the $R$ matrix we obtain the associate
hamiltonian that corresponds to a multi-state spin chain with no-isotropic
interaction. In that sense, we say that the models are a generalization to
$sl(n)$ of the $sl(2)$  XXZ  model.

In the third section, the eigenvalues of the transfer matrix of such models are
calculated with the YBE. The nested algebraic Bethe ansatz is used [11]. For
this calculation, we follow the method exposed in ref. [12]. A  detailed
account of the commutation rules derived from the YBE in every step of the
development is given. The results are obtained for the eigenvalues of the
transfer matrix in the ${U}_{q}\bigl(\widehat{sl(n)}\bigr)$, $n=2, 3 $  cases
and the spectrum of energy is calculated  for these cases and generalized to
$n>3$.

\chapter{Formulation}

Let   $sl(n)$ be the semisimple Lie algebra with generators ${e}_{i}, {f}_{i},
{h}_{i}, (i=1, \dots, n-1) $  and ${A}_{n-1}=({a}_{i,j})$ its Cartan matrix.
One fundamental representation with dimension $n$ is given by
$$\eqalignno{
&{e}_{i}\rightarrow{e}_{i,i+1},&(2.1a )\cr
&{f}_{i}\rightarrow{e}_{i+1,i} ,&(2.1b)\cr
&{h}_{i}\rightarrow{e}_{i,i} - {e}_{i+1,i+1},&(2.1c)\cr
}$$
where
$${({e}_{p,q})}_{i,j} = {\delta}_{p,i } {\delta}_{q,j},\qquad (
p,q,i,j=1,\cdots,n).\eqno(2.2)$$
Following the Drinfeld-Jimbo method  [2,8], the q-deformation of
${U}_{q}\bigl(sl(2)\bigr) $ is generated by
$${e}_{i},{f}_{i},{k}_{i},{k}_{i}^{-1},\qquad(i=1,\cdots,n-1)\eqno(2.3)$$
where  ${k}_{i}={q}^{{h}_{i}/2}$. These satisfy the following relations
$$\eqalignno{
&[{k}_{i},{k}_{j}]= 0,& (2.4a)\cr
&k_i e_j k_{i}^{-1}=q^{{a_{i,j}}/2} e_j, &(2.4b) \cr
&k_i f_j k_{i}^{-1}=q^{{-a_{i,j}}/2} f_j, &(2.4c) \cr
&[e_i,f_j]={\delta}_{i,j} {k_i^2-k_i^{-2}\over q-q^{-1}},&(2.4d)\cr
&k_i k_i^{-1}=k_i^{-1} k_i=1,&(2.4e)\cr
&\sum_{\nu=0}^{1-a_{i,j}}{(-1)^\nu  {1-a_{i,j} \brack \nu}_q e_i^{1-
a_{i,j}-\nu}  e_j e_i^\nu}=0,\qquad(i\neq j),&(2.4f)\cr
&\sum_{\nu=0}^{1-a_{i,j}}{(-1)^\nu  {1-a_{i,j} \brack \nu}_q f_i^{1-
a_{i,j}-\nu}  f_j f_i^\nu}=0,\qquad(i\neq j),&(2.4g)\cr
}
$$
where
$$\eqalignno{
&{n \brack j}={{[n]_q [n-1]_q\cdots[n-j+1]_q}\over [j]_q [j-1]_q\cdots [1]_q},
\qquad {n \brack 0}= 1,\qquad j\in N,&(2.5a)\cr
&[n]_q={q^n -q^{-n}\over q-q^{-1}}.&(2.5b)\cr}$$

When $q$ is not a root of the unity, an irreducible representation of
${U}_{q}\bigl(sl(n)\bigr) $ with dimension $n$ is obtained by using the
fundamental representation of $sl(n)$ given in (2.1). It is defined with the
substitutions
$$\eqalignno{
&{e}_{p}\rightarrow{e}_{p,p+1},&(2.6a )\cr
&{f}_{p}\rightarrow{e}_{p+1,p}, &(2.6b)\cr
&{k}_{p}\rightarrow q^{h_p /2} .&(2.6c)\cr
}$$

Besides ${U}_{q}\bigl(sl(n)\bigr) $ is a Hopf algebra with a coproduct
$\triangle$ defined with the elements of the base
$$\eqalignno{
&\triangle (e_i)=k_i \otimes e_i + e_i \otimes k_i^{-1},&(2.7a)\cr
&\triangle (f_i)=k_i \otimes f_i + f_i \otimes k_i^{-1}, &(2.7b)\cr
&\triangle (k_i^{\pm 1})=k_i ^{\pm 1}\otimes k_i ^{\pm 1}. &(2.7c)\cr
}$$
As this co-product is not commutative, we can as well define a new co-product
$\triangle^\prime$ given by
$$\eqalignno{
&{\triangle}^{\prime}(e_i)=k_i^{-1} \otimes e_i + e_i \otimes k_i ,&(2.8a)\cr
&{\triangle}^{\prime}(f_i)=k_i ^{-1}\otimes f_i + f_i \otimes k_i, &(2.8b)\cr
&{\triangle}^{\prime}(k_i^{\pm 1})=k_i ^{\pm 1}\otimes k_i ^{\pm 1}. &(2.8c)\cr
}$$
Both co-products are related by a transformation $R$ such that
$$
R  \cdot \triangle(a) = \triangle^{\prime} (a)  \cdot R ,\qquad \forall a \in
U_q (\widehat{sl(2)} ),\eqno (2.9)
$$
and $R$ verifies the Yang-Baxter equation
$$ R_{1,2} \cdot R_{1,3}  \cdot  R_{2,3} =  R_{2,3}  \cdot R_{1,3}  \cdot
R_{1,2}. \eqno(2.10)$$

The models that we are going to describe, will be derived from a representation
of the affine algebra $ {U}_{q}(\widehat{sl(n)})$, with Cartan matrix
${A}^{(1)}_{n-1}$. The generators  ${\left\{ {E}_{i}, {F}_{i}, {H}_{i}
\right\}}_{i=0}^{n-1}$ in a representation of such algebra can be obtained from
another representation of $U_{q} \left( sl\left( n \right) \right)$ using an
affine parameter $x$ as follows
$$
\eqalignno{
&E_i = x e_i , &(2.11a) \cr
&F_i = x^{-1} f_i ,&(2.11b) \cr
&H_i= h_i, \qquad (i=1,2,\cdots,n-1 ),&(2.11c)\cr
}$$
for the ordinary generators, and
$$
\eqalignno{
&E_0 =  x  [ f_{n-1} , f_1 ], &(2.12a) \cr
&F_0 =  x^{-1}  [ e_1  , e_{n-1} ], &(2.12b) \cr
&H_0= -h_1 - h_2 - \cdots -h_{n-1}, &(2.12c)\cr
}$$
for the affine generators.

The product of representations must obey
$$
R(x,y) \cdot \triangle_{(n, x) \otimes (n, y)} (a) = \triangle_{(n, x) \otimes
(n, y)} ^{\prime} (a) \cdot R(x, y) , \qquad \forall a \in
U_q\widehat{(sl(n))} ,\eqno (2.13 )
$$
where $(n, x)$ means a representation with dimension $n$ and affine parameter
$x$.

In this paper, one of our goals is to find the $R(x, y) $ matrix associated to
the fundamental  \break $n$-dimensional representations of  $
U_q\widehat{(sl(n))} $ when $q$ is not a root of the unity.

We begin with $ U_q\widehat{(sl(2))}$, where the two-dimensional
representation, for affine parameter $x$, is given by
 $$ \eqalignno{
&E_1 = \pmatrix{
0 & x \cr
0 & 0 \cr
},\qquad
F_1 = \pmatrix{
0 & 0 \cr
x^{-1} & 0 \cr
},\qquad
K_1=
\pmatrix{
q^{1/2} & 0 \cr
0 & q^{-1/2}  \cr
},&(2.14a)\cr
&E_0 = \pmatrix{
0 & 0 \cr
x & 0 \cr
},\qquad
F_0 = \pmatrix{
0 & x^{-1}  \cr
0 & 0 \cr
},\qquad
K_0=
\pmatrix{
q^{-1/2} & 0 \cr
0 & q^{1/2}  \cr
}.&(2.14b)
}
$$
We take now
$$
R_{sl(2)}=\left (\matrix{
a & 0 & 0 & 0 \cr
0 & b & c & 0 \cr
0 & c & b & 0 \cr
0 & 0 & 0 & a \cr
}\right ), \eqno(2.15)
$$
where $a$, $b$ and $c$ depend  on the parameters $x$ and $y$ that characterize
each representation involved in product representation $(n,x)\otimes(n,y)$.

If we impose the fulfilment of (2.13), we find
$$
a = y^2 -x^2 q^2, \qquad b = q ( y^2 - x^2 ),\qquad c =x y ( 1 - q^2),
\eqno(2.16)
$$
with what, we can write the $R$ matrix in the base ${(e_{p,r})}_{i,j} =
\delta_{p,i} \delta_{r,j} $,
$$
\eqalign{R^{sl(2)}(x,y) = &( y^2 -x^2 q^2) \sum_{i=1}^{2}{e_{i,i}\otimes
e_{i,i} } +
 q ( y^2 - x^2 ) \sum_{i,j=1 \atop i \neq j}^{2}{e_{i,i}\otimes e_{j,j}}  \cr
&+(1-q^2)  \sum_{i,j=1 \atop i < j}^{2}{(x^{j-i} y^{i-j }y^2 e_{i,j}\otimes
e_{j,i}+
x^{i-j} y^{j-i} x^2 e_{j,i}\otimes e_{i,j})}. \cr} \eqno (2.17)
$$

In the $U_q \widehat {(sl(3))} $ case, we can follow the same steps as before.
With (2.11), (2.12) and (2.6), we build the generators and  impose a matrix $R$
of the form
$$
 R^{sl(3)} =
\left (\matrix{
a & 0 & 0 & 0 & 0 & 0 & 0 & 0 & 0 \cr
0 & b & 0 & c & 0 & 0 & 0 & 0 & 0 \cr
0 & 0 & b & 0 & 0 & 0 & d & 0 & 0 \cr
0 & d & 0 & b & 0 & 0 & 0 & 0 & 0 \cr
0 & 0 & 0 & 0 & a & 0 & 0 & 0 & 0 \cr
0 & 0 & 0 & 0 & 0 & b & 0 & c & 0 \cr
0 & 0 & c & 0 & 0 & 0 & b & 0 & 0 \cr
0 & 0 & 0 & 0 & 0 & d & 0 & b & 0 \cr
0 & 0 & 0 & 0 & 0 & 0 & 0 & 0 & a \cr
}\right ).
\eqno(2.18)$$

  The fulfilment of the equation (2.13) needs that
$$
a = y^3 -x^3 q^2, \qquad b = q ( y^3 - x^3 ),\qquad c =x y^2 ( 1 - q^2),
\qquad d =x^2 y ( 1 - q^2), \eqno(2.19)
$$
that allow to express
$$
\eqalign{R^{sl(3)}(x,y) = &( y^3 -x^3 q^2) \sum_{i=1}^{3}{e_{i,i}\otimes
e_{i,i} } +
 q ( y^3 - x^3 ) \sum_{i,j=1 \atop i \neq j}^{3}{e_{i,i}\otimes e_{j,j}}  \cr
&+(1-q^2)  \sum_{i,j=1 \atop i < j}^{3}{(x^{j-i} y^{i-j }y^3 e_{i,j}\otimes
e_{j,i}+
x^{i-j} y^{j-i} x^3 e_{j,i}\otimes e_{i,j})}, \cr} \eqno (2.20)
$$
The generalization to $U_q \widehat {(sl(n))} $ is now obvious
$$
\eqalign{R^{sl(n)}(x,y) = &( y^n -x^n q^2) \sum_{i=1}^{n}{e_{i,i}\otimes
e_{i,i} } +
 q ( y^n - x^n ) \sum_{i,j=1 \atop i \neq j}^{n}{e_{i,i}\otimes e_{j,j}}  \cr
&+(1-q^2)  \sum_{i,j=1 \atop i < j}^{n}{(x^{j-i} y^{i-j }y^n e_{i,j}\otimes
e_{j,i}+
x^{i-j} y^{j-i} x^n e_{j,i}\otimes e_{i,j})} .\cr} \eqno (2.21)
$$
Taking into account that the YBE is homogeneous, we can substitute
$$
{y \over x} = \exp{(u)} ,\qquad q = \exp{(-\gamma)}, \eqno (2.22)
$$
in eq. (2.21) and write the $R$-matrix in the form
$$
\eqalign{
R^{sl(n)}(u) =& \sinh{( {n \over 2} u +\gamma) }  \sum_{i=1}^{n}{
e_{i,i}\otimes e_{i,i} }+
\sinh{({n \over 2}  u)} \sum_{i,j=1 \atop i \neq j}^{n}{ e_{i,i}\otimes e_{j,j}
}    \cr &+
\sinh{(\gamma)}  \sum_{i,j=1 \atop i \neq j}^{n} {    \exp{[(-j+i-{n \over 2}
\sign {(i-j)}) u]}
 e_{i,j}\otimes e_{j,i}   }   ,\cr} \eqno(2.23)
$$
where $u$ is the so called spectral parameter.

Associated to every solution of the YBE, we can find a solvable model. So we
introduce an one-dimensional lattice with a  $V_r=\bf C^n$ in every site $r$
and an auxiliary vector space $A= \bf C^n$. Now, we define an operator per site
$L_r (u) $,  acting on $V_r$ and $A$ and depending on the spectral parameter
$u$ and a new operator $R(u)$ acting on $A\otimes A$. Then, the YBE (2.10) can
be written in the usual form
$$
R(u-v) \cdot \left( L_r (u) \otimes L_r (v) \right) = \left( L_r (v) \otimes
L_r (u)  \right)\cdot R( u-v ), \eqno (2.24)
$$
where the $\otimes$ product is in the site space and $\cdot$ product is in $A
\otimes A$ tensorial space. With this new form for YBE, the expression for $L$
is the same as for $R$ in (2.23) and the new $R$ is obtained from (2.23) by
interchanging the indices $j$ and $m$ in every product  $e_{i,j}\otimes
e_{l,m}$ and  in his coefficient. Then we have
$$
\eqalign{
R (u) =& \sinh{( {n \over 2} u +\gamma) }  \sum_{i=1}^{n}{ e_{i,i}\otimes
e_{i,i} }+
\sinh{({n \over 2}  u)} \sum_{i,j=1 \atop i \neq j}^{n}{ e_{i,i}\otimes e_{j,j}
}    \cr &+
\sinh{(\gamma)}  \sum_{i,j=1 \atop i \neq j}^{n} {    exp{[(j-i-{n \over 2}
\sign {(j-i)}) u]}
 e_{i,i}\otimes e_{j,j}   },\cr} \eqno(2.25a)
$$
$$
\eqalign{
L_r (u) =& \sinh{( {n \over 2} u +\gamma) }  \sum_{i=1}^{n}{ e_{i,i}\otimes
e_{i,i} ^r}+
\sinh{({n \over 2}  u)} \sum_{i,j=1 \atop i \neq j}^{n}{ e_{i,i}\otimes
e_{j,j}^r }    \cr &+
\sinh{(\gamma)}  \sum_{i,j=1 \atop i \neq j}^{n} {    exp{[(j-i-{n \over 2}
\sign {(j-i)}) u]}
 e_{i,j}\otimes e_{j,i}^r   }.\cr} \eqno(2.25b)
$$

Since $L_r$ is defined in $A \otimes V_r \sim  \bf C^n \otimes \bf C^n$, we say
that it is a $n$-component solvable model, and it is generalization of the
model XXZ, defined in $ \bf C^2 \otimes \bf C^2$ with the Lie algebra $sl(2)$,
to $sl(n)$.

To obtain the associated hamiltonian, we should proceed in the same way that in
the $sl(2)$ case. We build the monodromy operator
$$
T(u) = \prod_{i=N}^{1} {L_i (u)} = L_N (u) \cdot L_{(N-1)} \cdots L_1 (u),
\eqno(2.26) $$
where the $\cdot$ product is understood as before in the auxiliary space and
$N$ is the length of the chain.

Taking
$$
F(u)= \trace_{aux.} ( T(u)), \eqno (2.27)
$$
the hamiltonian associated to the multi-state  $sl(n)$  chain is obtained when
we evaluate in $u=0$, the derivative with respect to $u$ of the logarithm of
$F$ . So, as in the $sl(2)$ case, the hamiltonian is defined by
$$
H = {2 \over n} \sinh{\gamma} {d \over du}{ \ln (F(u))} \big | _{u=0} - {N
\over n} \cosh{\gamma},  \eqno(2.28)
$$
that can be expressed as
$$
H = \sum_{r=1}^{N-1}{h_{r,r+1}},  \eqno (2.29)
$$
with
$$
\eqalign{
h_{r,r+1} = & \sum_{i,j=1 \atop i \neq j}^{n}{ e_{i,j}^{r} \otimes
e_{j,i}^{r+1}}+
{n-1 \over n} \cosh{(\gamma)}    \sum_{i=1}^{n}{ e_{i,i}^{r} \otimes
e_{i,i}^{r+1} }  \cr & +
\sum_{i,j=1 \atop i \neq j}^{n}{ \Bigl(  \bigl( {2 (j-i) \over n} - \sign
(j-i)\bigr) \sinh{( \gamma )} - {\cosh{( \gamma )} \over n}\Bigr)  e_{i,i}^{r}
\otimes e_{j,j}^{r+1} } .\cr}
\eqno (2.30)
$$
%
%
%
If we specify for $n = 2$, we obtain the hamiltonian corresponding to the XXZ
model
$$
H^{sl(2)}= {1 \over 2} \sum_{n=1}^{N}{ (\sigma _n^x \sigma_{n+1}^x +\sigma _n^y
\sigma_{n+1}^y +\cosh{(\gamma)}\sigma _n^z \sigma_{n+1}^z) },\eqno(2.31)
$$
where $\sigma^x, \sigma^y$  and $\sigma^z$ are the Pauli matrices. That model
has a well known solution.

For $n=3$ the hamiltonian obtained is
$$
H^{sl(3)}= {1 \over 2} \sum_{n=1}^{N}{\left( J_9 (\lambda_n^8 \lambda_{n+1}^3
+\lambda_n^3 \lambda_{n+1}^8 )+\sum_{\alpha =1}^{8}{\left( J_\alpha
\lambda_n^\alpha \lambda_n^\alpha  \right)} \right)} ,\eqno (2.32)
$$
with
$$J_1=J_2=J_4=J_5=J_6=J_7=1,\qquad J_3=J_8=\cosh{(\gamma)},\qquad J_9={
\sinh{(\gamma)}\over \sqrt{3} }, \eqno(2.34)
$$
where $ \lambda^\alpha  (\alpha=1,\cdots, 8)$ are the Gell-Mann matrices

%
\chapter{Solutions in the $sl(2)$ and $sl(3)$ cases}
The method that we are going to use is known as nested algebraic Bethe ansatz.
This method is a generalization to $n$ components of the normal algebraic Bethe
ansatz in two dimensions proposed by Fadeev and Takhtajan [13]. We will follow
in our calculation  the steps such as they are described in ref. [12].

We start  with  $sl(2)$ case or XXZ model. For this model the matrices $R$ and
$L$ are
$$
R(u,\gamma)=\left (\matrix{
a & 0 & 0 & 0 \cr
0 & c & b & 0 \cr
0 & b & c & 0 \cr
0 & 0 & 0 & a \cr
}\right ),\qquad
L(u,\gamma)=\left (\matrix{
a & 0 & 0 & 0 \cr
0 & b & c & 0 \cr
0 & c & b & 0 \cr
0 & 0 & 0 & a \cr
}\right ),\eqno(3.1)
$$
with
$$
\eqalignno{
&a=\sinh{(u+\gamma)},&(3.2a )\cr
&b= \sinh{(u)},&(3.2b)\cr
&c=sinh{(\gamma)}.&(3.2c)\cr
}$$
The solution to this model can be found in many places, in particular in ref.
[12] . We look for solutions of the eigenvalue equation for the operator $F(u)$
defined in (2.27). They are
$$\Lambda(u,{u}_{1},\cdots,{u}_{r})={\sinh{(u+\gamma)}}^{N}
\prod_{j}^{r}{\sinh{({u}_{j}-u+\gamma)} \over \sinh{({u}_{j}-u )} }
+{\sinh{(u)}}^{N} \prod_{j}^{r}{\sinh{(u-{u}_{j}+\gamma)} \over
\sinh{(u-{u}_{j})} }, \eqno(3.3)
$$
$r$ being  the number of sites with spin down in a lattice with length $N$. The
parameters ${u}_{j}, j=1,\cdots, r $, are the solutions of Bethe equations
$${\sinh{({u}_{k}+\gamma)} \over \sinh{({u}_{k})}}=
\prod_{l \neq k \atop l=1}^{r}
{{\sinh{({u}_{l}-{u}_{k}-\gamma)} \over \sinh{({u}_{l}-{u}_{k}+\gamma)}}},
\qquad
k=1,\cdots,r.
\eqno(3.4)
$$
The eigenvalues of the hamiltonian are obtained with the substitution in (2.28)
of $F(u)$ by $\Lambda(u)$.

In the $sl(3)$ case we have the $R$ and $L$ matrices
$$
R(u)=
\left (\matrix{
1 & 0 & 0 & 0 & 0 & 0 & 0 & 0 & 0 \cr
0 & d & 0 & b & 0 & 0 & 0 & 0 & 0 \cr
0 & 0 & c & 0 & 0 & 0 & b & 0 & 0 \cr
0 & b & 0 & c & 0 & 0 & 0 & 0 & 0 \cr
0 & 0 & 0 & 0 & 1 & 0 & 0 & 0 & 0 \cr
0 & 0 & 0 & 0 & 0 & d & 0 & b & 0 \cr
0 & 0 & b & 0 & 0 & 0 & d & 0 & 0 \cr
0 & 0 & 0 & 0 & 0 & b & 0 & c & 0 \cr
0 & 0 & 0 & 0 & 0 & 0 & 0 & 0 & 1 \cr
}\right ),\qquad
L(u)=
\left (\matrix{
1 & 0 & 0 & 0 & 0 & 0 & 0 & 0 & 0 \cr
0 & b & 0 & c & 0 & 0 & 0 & 0 & 0 \cr
0 & 0 & b & 0 & 0 & 0 & d & 0 & 0 \cr
0 & d & 0 & b & 0 & 0 & 0 & 0 & 0 \cr
0 & 0 & 0 & 0 & 1 & 0 & 0 & 0 & 0 \cr
0 & 0 & 0 & 0 & 0 & b & 0 & c & 0 \cr
0 & 0 & c & 0 & 0 & 0 & b & 0 & 0 \cr
0 & 0 & 0 & 0 & 0 & d & 0 & b & 0 \cr
0 & 0 & 0 & 0 & 0 & 0 & 0 & 0 & 1 \cr
}\right ),
\eqno (3.5)
$$
where we have taken into account that the YBE is homogeneous, then we have
redefined
$$
\eqalignno{
&b(u)= {\sinh{({3 \over 2} u)} \over \sinh{({3 \over 2} u+\gamma)} },&(3.6a)
\cr
&c(u)={\sinh{\gamma} \over \sinh{({3 \over 2} u+\gamma)} }{e}^{{ u\over
2}},&(3.6b) \cr
&d(u)={\sinh{\gamma} \over \sinh{({3 \over 2} u+\gamma)} }{e}^{{ -u\over
2}}.&(3.6c)
\cr
}$$
The monodromy matrix $T$ in (2.26) is specified as
$$
T(u)=\pmatrix{
A(u)& {B}_{2} (u) & {B}_{3}(u) \cr
{C}_{2} (u) & {D}_{2 2} (u) & {D}_{2 3}(u) \cr
{C}_{3} (u) & {D}_{3 2} (u) & {D}_{3 3}(u) \cr
}
\eqno (3.7)
$$
that is a matrix in the auxiliary space whose elements are operators on the
sites of the chain.
In order to solve the model we must apply the Bethe ansatz twice; in every step
we must introduce a set of parameters on which the eigenvectors and the
eigenvalues depend.

In the first step, inserting (3.7) in the YBE we find, with the usual notation,
$$
\eqalignno{
&B(u) \otimes B(v)= {R}^{(2)} (u-v) \cdot\bigl( B(v)\otimes B(u) \bigr) =
\bigl( B(v)\otimes B(u) \bigr)\cdot  {R}^{(2)} (u-v), & (3.8a) \cr
&A(u) B(v)=g(v-u) B(v)  A(u) -B(u) A(v) \cdot {\tilde{r}}^{(2)}(v-u),& (3.8b)
\cr
&D(u) \otimes B(v)=g(u-v) B(v) \otimes (D(u) \cdot{R}^{(2)} (u-v) )- B(u)
\otimes  ({r}^{(2)}(u-v) \cdot D(v) ),\cr & & (3.8c) \cr
}
$$
where
$$
{R}^{(2)}(u) =
\left (\matrix{
1 & 0 & 0 & 0 \cr
0 & d & b & 0 \cr
0 & b & c & 0 \cr
0 & 0 & 0 & 1 \cr
}\right ) ,\qquad
{r}^{(2)}(u)=\pmatrix{
{h}_{-}& 0 \cr
0 & {h}_{+}\cr
},\qquad
{\tilde{r}}^{(2)}(u)=
\pmatrix{
{h}_{+} & 0 \cr
0 & {h}_{-} \cr
}, \eqno (3.9)
$$
and $g(u)=1/b(u), {h}_{+}(u)= c(u)/ b(u)  $ and ${h}_{-}(u)= d(u)/ b(u)$.

Now, with the help of the relations (3.8), we look for solutions of
the equation
$$
F(u) \Psi ({\mu}_{1},\cdots,{\mu}_{r})=\Lambda(u, {\mu}_{1},\cdots,{\mu}_{r})
\Psi ({\mu}_{1},\cdots,{\mu}_{r}),\eqno (3.10)
$$
of the form
$$
\Psi(\vec{\mu})=\Psi ({\mu}_{1},\cdots,{\mu}_{r})={X}_{{i}_{1},\cdots,{i}_{r}}
{B}_{{i}_{1}}({\mu}_{1})\otimes \cdots \otimes{B}_{{i}_{r}}({\mu}_{r})
\parallel 1>, \eqno (3.11)
$$
being
$$
 \parallel 1>={
\left (\matrix{
1\cr
0\cr
0\cr
}\right )}_{1} \otimes \cdots \otimes
{
\left (\matrix{
1\cr
0\cr
0\cr
}\right )}_{N} .\eqno(3.12)
$$
This condition introduces the first set of parameters ${\left\{ {\mu}_{j}
\right\}}^{r}_{j=1}$.

To begin with, since $\parallel 1>$ is eigenvector of  $A(u)$ and ${D}_{i,i}$
with eigenvalues  $1$ and $b(u)^{N} \delta_{i,j}$ respectively, we apply these
operators on $\Psi$ and, by using the commutation relations (3.8), we push the
operators $A$ or $D_{i,j}$ through the $B$ to the right . When either $A$ or
$D$ reaches $\parallel 1> $ they reproduce this vector again. Since the
commutation relations have two terms, this procedure generates a lot of terms.
Some of them have the same order of the arguments in the $B$ product; we call
them wanted terms. The others have some $B({\mu}_{j})$ replaced by $B(u)$ and
we call them unwanted terms.

When we apply  $F= A+D_{2,2} +D_{3,3}$  to $ \Psi
({\mu}_{1},\cdots,{\mu}_{r})$, we collect the unwanted terms and require them
to have a vanishing sum. This condition gives us a set of equations for the
parameters. The sum of the wanted terms will be required to be proportional to
$\Psi$, providing us with the second part of equation (3.10).

So, the application of $ A(u)$ to $\Psi$, gives the wanted term
$$
\prod_{j=1}^{r}{g({\mu}_{j} - u)} {B}_{{j}_{1}}({\mu}_{1})\cdots
{B}_{{j}_{r}}({\mu}_{r})   {X}_{{j}_{1}, \cdots, {j}_{r}} \parallel 1>,
\eqno(3.13)
$$
and the $k-th$ unwanted term
$$
  \prod_{j=1 \atop j\neq k}^{r}{g({\mu}_{j} - {\mu}_{k})}
\left( B(u) \tilde{r}^{(2)}({\mu}_{k} - u)  \right)
\otimes {B}({\mu}_{k+1})\otimes \cdots \otimes {B}({\mu}_{r})\otimes
{B}({\mu}_{1})   \otimes {B}({\mu}_{k-1}) {M}^{(k-1)} X \parallel
1>,\eqno(3.14)
$$
${M}^{(k-1)} $ being the tensor obtained from the commutation relations (3.8a),
which produces a cyclic permutation
$$
{B}({\mu}_{1})\otimes \cdots \otimes {B}({\mu}_{r})={B}({\mu}_{k+1})\otimes
\cdots {B}({\mu}_{r})\otimes {B}({\mu}_{1}) \cdots  \otimes {B}({\mu}_{k-1})
{M}^{(k-1)}, \eqno (3.15)
$$

In the same form, the application of ${D}_{2,2}+{D}_{3,3}$ to $\Psi$ produces
the wanted term
$$
{\left( {1 \over g\left( {\mu}_{l} \right)} \right)}^{N}\prod_{j=1}^{r}{g(u
-{\mu}_{j} )} {B}({\mu}_{1})\otimes\cdots  \otimes{B}({\mu}_{r}) {F}_{r}^{(2)}
\left( u,\vec{\mu} \right)  X \parallel 1>,  \eqno(3.16)
$$
and the $k$-th unwanted term
$$   \eqalign {
{\left( {1 \over g\left( {\mu}_{k} \right)} \right)}^{N}
& \prod_{j=1 \atop j\neq k}^{r}{g({\mu}_{j} - {\mu}_{k})}
\left( B(u) \tilde{r}^{(2)}({\mu}_{k} - u)  \right)
\otimes {B}({\mu}_{k+1})
\otimes \cdots \cr  \cdots &\otimes {B}({\mu}_{r})\otimes {B}({\mu}_{1})
\otimes \cdots \otimes{B}({\mu}_{k-1}) F_{r}^{(2)}
({\mu}_{k},\vec{\mu}){M}^{(k-1)} X \parallel 1>,  \cr}
\eqno (3.17)
$$
the operator
$$F_{r}^{(2)}=\sum_{j=2,3}
{T_{r}^{\left( 2
\right)}(u,\vec{\mu})_{j,{j}_{1},\cdots,{j}_{r}}^{j,{i}_{1},\cdots,{i}_{r}}  },
\eqno(3.18)
$$
being the trace of a tensor acting on the index  of the $B$'s with the values
$i=2, 3$ and defined by
$$
T_{r}^{\left( 2
\right)}(u,\vec{\mu})_{j,{j}_{1},\cdots,{j}_{r}}^{i,{i}_{1},\cdots,{i}_{r}}=
{{R}^{\left( 2 \right)  }}_{  {l}_{1},{m}_{1}}^{i,{j}_{1}} (u-{\mu}_{1})
{{R}^{\left( 2 \right)  }}_{  {l}_{2},{m}_{2}}^{{m}_{1},{j}_{2}} (u-{\mu}_{2})
\cdots
{{R}^{\left( 2 \right)  }}_{  {l}_{r},j}^{ {m}_{r-1},{j}_{r}} (u-{\mu}_{r}),
 \eqno(3.19)
$$
In order to the sum of the wanted terms solve (3.10),
${X}_{{i}_{1},\cdots,{i}_{r}}$ must be eigenstate of $F_{r}^{(2)} $
$$
F_{r}^{(2)} (u,\vec{\mu}) X= {\Lambda}_{(2)}(u,\vec{\mu}) X, \eqno(3.20)
$$
However, the cancelation of  unwanted terms gives
$$
{\Lambda}_{(2)}({\mu}_{k},\vec{\mu}) =\left( g\left( {\mu}_{k} \right)
\right)^{N}
\prod_{j \ne k \atop j=1}^{r}{{g\left( {\mu}_{j}-{\mu}_{k} \right) \over
g\left( {\mu}_{k}-{\mu}_{j}\right) }},
\eqno (3.21)
$$
The second step then is to diagonalize the equation (3.20). We proceed
analogously to the first step. The operator $T_{r}^{\left( 2
\right)}(u,\vec{\mu})_{j,{j}_{1},\cdots,{j}_{r}}^{i,{i}_{1},\cdots,{i}_{r}}$ is
as a dimension two matrix in the index $i,j$  written as
$$T_{r}^{\left( 2 \right)}(u,\vec{\mu})=
\pmatrix{
A_{r}^{\left( 2 \right)}(u,\vec{\mu}) & B_{r}^{\left( 2
\right)}(u,\vec{\mu})\cr
C_{r}^{\left( 2 \right)}(u,\vec{\mu}) & D_{r}^{\left( 2 \right)}(u,\vec{\mu})
\cr
},
\eqno(3.22)
$$
It satisfies a YBE with the matrix $R^{(2)}$ in (3.9),
$$
R^{(2)}\left( u-v \right) \cdot \left(    T_{r}^{\left( 2 \right)}(u,\vec{\mu})
\otimes
T_{r}^{\left( 2 \right)}(v,\vec{\mu})\right)=
\left(    T_{r}^{\left( 2 \right)}(v,\vec{\mu}) \otimes
T_{r}^{\left( 2 \right)}(u,\vec{\mu})\right)\cdot R^{(2)}\left( u-v \right),
\eqno(3.23)
$$
that gives the relations
$$  \eqalignno {
&{A}^{(2)}(u) \cdot {B}^{(2)}(v)= g(v-u) {B}^{(2)}(v) \cdot {A}^{(2)}(u)-
{h}_{+}(v-u) {B}^{(2)}(u) \cdot {A}^{(2)}(v), &  (3.24a) \cr
&{D}^{(2)}(u) \cdot {B}^{(2)}(v)= g(u-v) {B}^{(2)}(v) \cdot {D}^{(2)}(u)-
{h}_{-}(u-v) {B}^{(2)}(u) \cdot {v}^{(2)}(v), &  (3.24b) \cr
&[ {B}^{(2) }(u) ,  {B}^{(2) }(v)] =0. &(3.24c)\cr}
$$
We take now
$$
{\parallel 1>}^{(2)}=\bigotimes_{i=1}^{r}{{\left (\matrix{
1\cr
0\cr
}\right )}_{i}}, \eqno (3.25)
$$
that is a eigenvector of $A_{(2)}$ and $D_{(2)}$
$$ \eqalignno{
&A^{(2)} \left( u, \vec{\mu} \right){\parallel 1>}^{(2)} = {\parallel
1>}^{(2)},&(3.26a)\cr
&D^{(2)} \left( u, \vec{\mu} \right){\parallel 1>}^{(2)} = \left(
\prod_{i=1}^{r}{{1 \over g(u-{\mu}_{i})}} \right){\parallel
1>}^{(2)},&(3.26b)\cr
&C^{(2)} \left( u, \vec{\mu} \right){\parallel 1>}^{(2)} =0, &(3.26c)
}
$$
and, as in the first step, we look for eigenvectors of the form
$${\Psi}^{(2)}={B}^{(2)}({\lambda}_{1},\vec{\mu} ) \cdots
{B}^{(2)}({\lambda}_{s},\vec{\mu} ) {\parallel 1>}^{(2)}, \eqno (3.27)
$$
that introduce the dependence of the eigenvalues on a new set of parameters
$\left\{ {\lambda}_{i} \right\}_{i=1}^{s}$.

We follow the same procedures as in the first step, but now in two dimensions,
and we find that the wanted terms give the eigenvalues
$$
{\Lambda}_{(2)}(u,\vec{\mu} ,\vec{\lambda} )=
\prod_{i=1}^{s}{g({\lambda}_{i}-u)}+\prod_{i=1}^{s}{g(u-{\lambda}_{i})}
\prod_{j=1}^{r}{{1 \over g(u-{\mu}_{j})}},
\eqno (3.28)
$$
whereas the cancelation of the unwanted terms imposes the equations
$$
{h}_{+}({\lambda}_{k}-u) \prod_{i=1 \atop i\ne
k}^{s}{g({\lambda}_{i}-{\lambda}_{k})}+
{h}_{-}(u-{\lambda}_{k}) \prod_{i=1\atop i \ne
k}^{s}{g({\lambda}_{k}-{\lambda}_{i})}
\prod_{j=1}^{r}{{1 \over g({\lambda}_{k}-{\mu}_{j})}} =0.
\eqno (3.29)
$$
As  ${h}_{+}(x)+{h}_{-}(-x)=0$,  the last relations give the equations for the
$\left\{ {\lambda}_{i} \right\}_{i=1}^{s}$ parameters
$$
\prod_{j=1}^{r}{g({\lambda}_{k}-{\mu}_{j})}=
\prod_{i=1 \atop i \ne  k}^{s}{{ g({\lambda}_{k}-{\lambda}_{j})\over
g({\lambda}_{j}-{\lambda}_{k})}}, \qquad k=1,\cdots,s
\eqno(3.30)
$$
and, due to ${1 \over g(0) }=0$, the eigenvalues $\Lambda(u,\vec{\mu} )$ in
(3.10), for $ u={\mu}_{k}$, are
$$
\Lambda({\mu}_{k},\vec{\mu} )=\prod_{i=1}^{s}{g({\lambda}_{i}-{\mu}_{k})}.\eqno
(3.31)
$$
Then  our results are the eigenvalues of $F(u)$, given by the equations
$$\eqalignno{
\Lambda(u,\vec{\mu} ,\vec{\lambda} )=&
\prod_{i=1}^{r}{g({\mu}_{i}-u)} +{\left( {1 \over g(u)} \right)}^{N}
\prod_{i=1}^{r}{g(u-{\mu}_{i})} \Lambda_{(2)}(u,\vec{\mu} ,\vec{\lambda} ),
& (3.32a)\cr
\Lambda_{(2)}(u,\vec{\mu} ,\vec{\lambda} )=&
\prod_{i=1}^{s}{g({\lambda}_{i}-u)} +\prod_{i=1}^{s}{g(u-{\lambda}_{i})}
\prod_{j=1}^{r}{{1 \over g(u-{\mu}_{j})}}, &(3.32b)\cr
}
$$
and the parameters ${\left\{ {\mu}_{i} \right\}}_{i=1}^{r}$ and ${\left\{
{\lambda}_{i} \right\}}_{i=1}^{s}$, solutions of the coupled equations
$$\eqalignno{
\prod_{i=1}^{s}{g({\lambda}_{i}-{\mu}_{k})} =&
\left( g( {\mu}_{k}\right)^{N}
\prod_{j=1\atop j \ne k}^{r}{{g({\mu}_{j}-{\mu}_{k})\over
g({\mu}_{k}-{\mu}_{j})}},  &(3.33a)\cr
\prod_{j=1}^{r}{g({\lambda}_{k}-{\mu}_{j})} =&
\prod_{i=1\atop i \ne k}^{s}{{g({\lambda}_{k}-{\lambda}_{i})\over
g({\lambda}_{i}-{\lambda}_{k})}}. &(3.33b)
\cr}
$$
Every set of solutions for $1 \leq s \leq r\leq N $ of these coupled equations
determines an eigenvalue of $F$.

We can now substitute the function $g(x)$
$$
g(x)= {\sinh{({3 \over 2} x+ \gamma)} \over \sinh{({3 \over 2} x)}}, \eqno
(3.34)
$$
and use it to calculate the energy spectrum of the hamiltonian with (2.28).
Then, as ${1 \over g(0)} =0$, we have from (3.32a)
$$
\Lambda(0,\vec{\mu} ,\vec{\lambda} )=\prod_{i=1}^{r}{g({\mu}_{i})}, \eqno
(3.35)
$$
and
$$
{d \over du} \Lambda(u,\vec{\mu} ,\vec{\lambda} ) \big |_{u=0}=
\sum_{j=1}^{r}{{d \over du}[g({\mu}_{j}-u) ]\big |_{u=0}}
\prod_{i=1\atop i \ne
j}^{r}{g({\mu}_{i})}=\sum_{j=1}^{r}{f({\mu}_{j})\prod_{i=1\atop i \ne
j}^{r}{g({\mu}_{i})} },\eqno (3.36)
$$
with
$$
f\left( x \right)=-{3 \over 2} {\sinh{( \gamma) } \over \sinh{({3 \over 2} x)}
}.
\eqno (3.37)
$$
Then, the energy is
$$
E =
{3 \over 2} \sinh{(\gamma)}\sum_{j=1}^{r}{\left[ \sinh{({3 \over 2}
{\mu}_{j}+\gamma)}
\sinh{({3 \over 2} {\mu}_{j})} \right]^{-1}} \eqno (3.38)
$$
As we can see in the last expression, the energy only depends on the first set
of parameters $\left\{ {\mu}_{j} \right\}_{j=1}^{r}$.

As conclusion, we have found a family of integrable models based in the
$U_q\left(\widehat{sl(n)} \right)$ quantum group.
The  $n=2$ it is the same as the six vertex model that can be found in ref
[12]. For n=3 we have a model described with a matrix whose elements are the
functions $b$, $c$ and $d$ given in (3.2)  and depending on $\gamma$ and the
spectral parameter $u$.  It is different of the models studied in ref. [12]
since its hamiltonian can be proved to be different.

The generalization to $n>3$ follows easily. The number of functions on which
depend the matrix elements of the model is increased. Analogously  to the $c$
and $d$ functions, now we will have other functions with different exponentials
in $\gamma$. The function $b$, as can be see from (2.25), will be
$$
b(u,\gamma)={\sinh{({n \over 2} u)} \over\sinh{({n \over 2} u+\gamma) }}, \eqno
(3.39)
$$
and the energy spectrum, for any value of $n$, depends only of the parameters
${\left\{ {\mu}_{j} \right\}}_{j=1}^{r}$ introduced in the first step. It turns
out
$$
E =
{n \over 2} \sinh{(\gamma)}\sum_{j=1}^{r}{\left[ \sinh{({n \over 2}
{\mu}_{j}+\gamma)}
\sinh{({n \over 2} {\mu}_{j})} \right]^{-1}}, \eqno (3.40)
$$
obtained from eq. (3.18) by the substitution of ${3 \over 2}$ by ${n \over 2}$.
Eq (3.40) is applicable also in the case $n=2$.

{\bf Acknowledgements}

We thank J. Sesma for a careful reading of the manuscript. This work was
partially supported by the Direcci\'{o}n General de Investigaci\'{o}n
Cient\'{\i}fica y T\'{e}cnica Grant No AEN 90-0030
\vfill
\eject

%
%
%
\centerline{\bf References}

\item{[1]}{B. Sutherland, J. Math. Phys. 11, (1970) 3183}
\item{[2]}{R. J. Baxter, Ann. Phys. 70, (1970) 323}
\item{[3]}{L. D. Faddeev and E. K. Sklyanin,  Sov. Phys. Dokl. 23, (1978) 900}
\item{[4]}{L. D. Faddeev, Sov. Sci. Rev. Math. Phys. C1, (1981) 107}
\item{[5]}{V.G. Drinfeld, Proceedings of the I. C. M. 1986, A.M. Gleason editor
(A.M.S.1987)}
\item{[6]}{A. Berkovich, C. G\'{o}mez and G. Sierra, Int. J. Mod. Phys. B 7,
(1992) 1939}
\item{[7]}{C. G\'{o}mez and G. Sierra, Phys. Lett. B 285, (1992) 126.}
\item{[8]}{V. Pasquier and H. Saleur, Nucl. Phys. B 330, (1990) 523}
\item{[9]}{N. Dagupta and A. R. Chowdhury, Jour. Phys. A 26, (1993) 5427}
\item{[10]}{M. Jimbo, Lett. Math. Phys.10, (1985) 63.\hfill\break
M. Jimbo, Lett. Math. Phys. 11, (1986) 247}
\item{[11]}{B. Sutherland, Phys. Rev. B 12,  (1975) 3795.\hfill\break
B. Sutherland, J. Math. Phys. 12, (1971) 246}
\item{[12]}{H.J. de Vega, Int. J. Mod. Phys. A 4, (1989) 2371}
\item{[13]}{L. D. Faddeev and L. Takhtajan, Russ. Math. Surveys 34, (1979) 11}

\end